\documentclass[a4paper,11pt]{article}
\usepackage{pos}

\title{Lattice Correlation Functions from Differential Equations}

\author*[a]{Federico Gasparotto}
\author[a,b]{Andreas Rapakoulias}
\author[a]{Stefan Weinzierl}
\author[a]{Xiaofeng Xu}

\affiliation[a]{PRISMA Cluster of Excellence, Institut für Physik,
Johannes Gutenberg-Universität Mainz, D-55099 Mainz, Germany}

\affiliation[b]{Institute for Theoretical Physics, University of Regensburg, 93040 Regensburg, Germany}

\vspace*{4cm}

\emailAdd{fgasparo@uni-mainz.de}
\emailAdd{andreas.rapakoulias@ur.de}
\emailAdd{weinzierl@uni-mainz.de}
\emailAdd{xiaxu@uni-mainz.de}

\abstract{We discuss how methods developed in the context of perturbation theory can be applied to the computation of lattice correlation functions, in particular in the non perturbative regime. The techniques we consider are integration-by-parts identities (supplemented with symmetry relations) and the method of differential equations, cast in the framework of twisted co-homology. We report on calculations of correlation functions for a scalar $\lambda \phi^4$ theory and lattices of small size, both in Euclidean and Minkowskian signature.}

\FullConference{16th International Symposium on Radiative Corrections: Applications of Quantum Field Theory to Phenomenology (
RADCOR2023)\\
28th May - 2nd June, 2023\\
Crieff, Scotland, UK\\}

\begin{document}
\maketitle

\section{Introduction}
Correlation functions are key objects in Quantum Field Theory (QFT) and the quest to compute them is of the utmost importance in Physics. Calculating correlation functions in full generality is often a prohibitive task, and we have to rely on different approximations. Traditionally, correlation functions are addressed in the context of perturbation theory as a series expansion with respect to a small coupling, where individual contributions are expressed in terms of Feynman diagrams. Alternatively, they are studied in the context of Lattice Field Theory, considering a discrete version of Euclidean space-time and benefiting from powerful Monte-Carlo methods for the integration. Following~\cite{Weinzierl:2020nhw} and~\cite{Gasparotto:2022mmp,Gasparotto:2023roh}, in this contribution we review how techniques and tools developed in the context of perturbation theory can be applied to the study of lattice correlation functions, primarily at the non perturbative level: integration-by-parts identities (IBPs)~\cite{Chetyrkin:1981qh,Tkachov:1981wb,Laporta:2000dsw} (supplemented with symmetry relations) and the method of differential equations (DEQs)~\cite{Kotikov:1990kg,Remiddi:1997ny,Gehrmann:1999as}, cast in the mathematical framework of twisted co-homology~\cite{Mastrolia:2018uzb}.
\section{Setup and Notation}
Let us consider for concreteness a scalar $\lambda \phi^4$ model, whose action in the continuum limit reads
\begin{equation}
    S = \int d^D x \frac{1}{2} \partial^{\mu} \phi(x) \partial_{\mu} \phi(x) - \frac{m^2}{2} \phi^2(x) - \lambda \phi^4(x), \qquad D \in \mathbb{N}.
\end{equation}
Correlation functions are defined as
\begin{equation}
 G_{N}(x_1, \dots, x_N) =  \frac{\int \mathcal{D} \phi \phi(x_1) \dots \phi(x_N) e^{i S}}{\int \mathcal{D} \phi e^{i S}}.
\label{eq:correlator_continuum}
\end{equation}
In the context of Lattice Field Theory, we consider a periodic lattice $\Lambda$ with spacing $a$ and $L_{\mu}$ points in the $\mu-$direction, such that the total number of lattice sites is $N=\prod_{\mu=0}^{D{-}1} L_{\mu}$. Compared to the continuum limit, we have the following identifications
\begin{equation}
    \int d^{D}x \rightsquigarrow a^D\sum_{x \in \Lambda}, \qquad \qquad \partial_{\mu} \phi(x) \rightsquigarrow \frac{\phi(x+a \hat{e}_{\mu})-\phi(x)}{a},
\label{eq:substitution_continuum_lattice}
\end{equation}
where $x+a\hat{e}_{\mu}$ represents the lattice point adjacent to $x$ in the $\mu-$direction. Employing a Wick rotation$-$in order to land on a space-time with Euclidean signature$-$and the substitutions in eq.~(\ref{eq:substitution_continuum_lattice}), the Euclidean action on the lattice reads
\begin{equation}
    S_E =  \sum_{x \in  \Lambda} \left[  {-} \sum_{\mu=0}^{D-1} \phi(x) \, \phi(x{+}a \hat{e}_{\mu}) {+} \left( D  {+} \frac{m^2}{2} \right) \phi^2(x) {+}\lambda \phi^4(x)  \right].
\end{equation}
Alternatively, without resorting to Wick rotation, we can work directly in Minkowskian signature$-$i.e. we consider ``real time" quantities$-$with an action given by
\begin{equation}
    S_M = {i} \sum_{x \in  \Lambda} \left[  \phi(x) \, \phi(x{+}a \hat{e}_{0})  {-} \sum_{\mu=1}^{D-1} \phi(x) \, \phi(x{+}a \hat{e}_{\mu}) {+} \left( D  {+} \frac{m^2}{2}{-}2 \right) \phi^2(x) {+}\lambda \phi^4(x)  \right].
\label{eq:action_Minkowskian}
\end{equation}
In either case ($E$ for Euclidean and $M$ for Minkowskian), the action has the following general structure
\begin{equation}
    \begin{split}
        S_{\bullet} & = \text{``polynomial in the fields $\phi(x_i)$''} \\
                    &  = S_{\bullet}^{\text{next neigh.}}+S_{\bullet}^{(2)}+\lambda S_{\bullet}^{(4)}, \qquad \qquad \qquad \bullet = E,M,
    \end{split}
\label{eq:action_general}
\end{equation}
where $S_{\bullet}^{\text{next neigh.}}$ constitutes the interaction terms among adjacent lattice points, $S_{\bullet}^{(2)}$ the sum of quadratic terms and $S_{\bullet}^{(4)}$ the sum of quartic terms.\\
~\\
Our goal is to develop a computational strategy for integrals of the following form
\begin{equation}
    I_{\nu_1 \dots\nu_N} = \int_{\mathbb{R}^N}\, \exp(-S_{\bullet})\, \Phi, \qquad \Phi =\phi^{\nu_1}(x_1) \dots \phi^{\nu_N}(x_N) \, d^N \phi, \qquad \bullet=E,M,
    \label{eq:integral_family}
\end{equation}
where the indices $\nu_i$ are non-negative integers.\\
Once integrals in eq.~(\ref{eq:integral_family}) are known, then correlation functions on a lattice are given by (cf. eq.~(\ref{eq:correlator_continuum}))
\begin{equation}
    G_{\nu_1 \dots \nu_N} = \frac{I_{\nu_1 \dots\nu_N}}{I_{0 \dots 0}}.
\end{equation}
Actually, for reasons that will become transparent later on, we will focus on a slightly more general class of integrals: we introduce an auxiliary parameter in the action, dubbed ``auxiliary flow'' and denoted by $t$, according to
\begin{equation}
    S_{\bullet} \rightsquigarrow S_{t,\bullet}= t \cdot  S_{\bullet}^{\text{next  neigh.}}+S_{\bullet}^{(2)}+\lambda S_{\bullet}^{(4)}, \qquad \bullet=E,M.
\label{eq:Flow_action}
\end{equation}
The case of our interest is the limit $t \to 1$, where eq.~(\ref{eq:Flow_action}) reduces to eq.~(\ref{eq:action_general}).\\

Let us conclude this preliminary section  emphasizing that, while Euclidean correlators can be computed via very efficient Monte-Carlo methods, Minkowskian correlators are more challenging due to the oscillating behaviour of the integrand (cf. eq.~(\ref{eq:action_Minkowskian})).  
\section{Twisted Co-Homology}
Readers familiar with multi-loop calculus will certainly realize that the, a priori, infinite set of integrals described by eq.~(\ref{eq:integral_family}) are not all independent: IBPs$-$loosely speaking the vanishing of a total differential under the integral sign$-$guarantee the existence of linear relations among them. The mathematical framework of twisted co-homology formalizes this idea. Let us consider an $(N-1)-$differential form $\xi$, simple algebraic manipulations show that
\begin{equation}
    0 = \int_{\mathbb{R}^N} d \left( \exp(-S_{t,\bullet}) \xi \right) = \int_{\mathbb{R}^N}  \exp(-S_{t,\bullet}) \nabla_{-dS_{t,\bullet}} \xi, \qquad \bullet=E,M,
\label{eq:IBP}
\end{equation}
where we introduced
\begin{equation}
  \nabla_{-dS_{t,\bullet}} \xi = d \xi - d S_{t,\bullet} \wedge \xi, \qquad \bullet=E,M.
\end{equation}
Eq.~(\ref{eq:IBP}) suggests that there is a huge redundancy while considering differential forms separately; we can group them into equivalence classes, declaring that two differential forms are in the same equivalence class if they differ by $\nabla_{-dS_{t,\bullet}} \xi$ for any $\xi$. Equivalence classes, denoted by $\langle \Phi|$ are elements of the twisted co-homology group $\operatorname{H}^N$. Mathematically we have
\begin{equation}
\langle \Phi | : \Phi \sim \Phi + \nabla_{-dS_{t,\bullet}} \xi, 
\qquad
\langle \Phi | \in \operatorname{H}^N = \frac{\text{Ker}\, \nabla_{-d S_{t,\bullet}}}{\text{Im}\, \nabla_{-d S_{t,\bullet}}}, \qquad \bullet=E,M.
\qquad
\end{equation}
\subsection{Basis of Differential Forms}
The twisted co-homology group is a finite dimensional space, and, in the case at hand, it admits a $3^N$-dimensional basis~\cite{Weinzierl:2020nhw}, whose representatives are chosen as
\begin{equation}
    \phi^{\nu_1}(x_1) \dots \phi^{\nu_N}(x_N), \qquad 0 \leq \nu_1, \dots, \nu_N \leq 2.
\end{equation}
This fact can be justified as follows. In the case of our interest, we can identify the following qualitative behaviour 
\begin{equation}
  \nabla_{-dS_{t,\bullet}} \propto  - d S_{t,\bullet}  = - c_{\bullet}\sum_{x \in \Lambda} \phi^3(x) \, d\phi(x)+\text{``lower degree''},
    \qquad \bullet=E,M,
\end{equation}
where $c_{\bullet}=4 \lambda$ in the Euclidean case (i.e. $\bullet=E$) or $c_{\bullet}=  4 i \lambda$ in the Minkowskian case (i.e. $\bullet=M$) and ``lower degree'' represents a linear combination of differential forms$-$monomials$-$of degree strictly lower than $3$, whose explicit expressions are not relevant for the discussion.\\ 
We can consider a certain differential $N$-form that is not in the basis, say for concreteness
\begin{equation}
    \Phi = \phi^{\nu_1}(x_1) \dots \phi^{\nu_k}(x_k) \dots \phi^{\nu_N}(x_N) d^N \phi, \qquad   \exists \, \nu_k >2;
\end{equation}
then we can always construct the corresponding $(N-1)$-form as\footnote{We employ the notation $d^{N-1}\phi= d \phi(x_1) \wedge \dots \wedge d\hat{ \phi}(x_k) \wedge \dots \wedge d \phi(x_N)$, where $d \phi(x_k)$ is omitted.}
\begin{equation}
    \xi_{\Phi} =\frac{(-1)^{k-1}}{c_{\bullet}} \phi^{\nu_1}(x_1) \dots \phi^{\nu_k{-}3}(x_k) \dots \phi^{\nu_N}(x_N) d^{N{-}1} \phi, \qquad \bullet=E,M.
\end{equation}
The key point is that we can systematically replace $\Phi$ with another representative within the same equivalence class, according to 
\begin{equation}
    \Phi \sim \Phi + \nabla_{-d S_{t,\bullet}} \xi_{\Phi} = \Phi - \Phi +\text{``lower degree"}, \qquad \bullet=E,M.
\label{eq:cohomology_calss}
\end{equation}
The leading terms in the r.h.s. in eq.~(\ref{eq:cohomology_calss}) cancel, and we are left with a linear combination of monomials whose degrees are strictly lower than the original one. By repeated application of eq.~(\ref{eq:cohomology_calss}) we land on a combination of basis elements.\\ Turning the argument around: given any $\Phi$ we can always construct a suitable IBP relation such that the original object is replaced with a linear combination of ``simpler'' ones$-$i.e. monomials of lower degree$-$:
\begin{equation}
    \nabla_{-d S_{t,\bullet}} \xi_{\Phi} =0 \,\, \Rightarrow \,\, \Phi \to \text{``lower degree"}, \qquad \bullet=E,M.
\label{eq_reduction_substitution_rule}
\end{equation}
Eq.~(\ref{eq_reduction_substitution_rule}) implies that the reduction onto basis elements is just a repeated substitution rule, which can be implemented straightforwardly in computer algebra systems\footnote{See also~\cite{Gasparotto:2023roh} for efficient implementations of the reduction algorithm.}.\\
We conclude this paragraph with an important observation: eq.~(\ref{eq:cohomology_calss}) (or equivalently eq.~(\ref{eq_reduction_substitution_rule})) always produces positive powers of the variable $t-$i.e. polynomials in $t-$in the r.h.s.

Let us consider simple lattices with $L=2$ points in each direction, for various dimensions $D$, such that the total number of lattice sites is $N=2^D$. The dimension of the corresponding co-homology group$-$i.e. the number of independent differential forms$-$is given by
\begin{equation}
    \begin{tabular}{ |c|c|c|c|c| } 
 \hline
 D & 1 & 2 & 3 & 4  \\ 
\text{$N$} & 2 & 4 & 8 & 16\\
\text{$\#$\,differential forms} & 9 & 81 & 6\,561 & 43\, 046\, 721\\
 \hline
\end{tabular}
\label{eq.size_cohomology}
\end{equation}
\subsection{Basis of Integrals}
At this point, it is important to stress that there is another source of redundancy in the problem at hand, namely symmetry relations. In order to appreciate this fact, let us consider the case of an Euclidean action in $D=1$ with two points; its explicit expression reads
\begin{equation}
    S_{t,E}=-2 t \phi(x_1) \phi(x_2) + \left( 1 +  \frac{m^2}{2} \right) (\phi^2(x_1)+\phi^2(x_2))+\lambda ( \phi^4(x_1)+\phi^4(x_2)). 
\label{eq:SE_D1}
\end{equation}
It is straightforward to see that eq.~(\ref{eq:SE_D1}) is invariant under the transformations
\begin{equation}
    \phi(x_i) \to - \phi(x_i), \qquad \phi(x_1) \leftrightharpoons \phi(x_2). 
\end{equation}
The former is a global $\mathbb{Z}_2$ symmetry, while the latter corresponds to a permutation of the lattice points $x_1$ and $x_2$. On the one hand, differential forms are blind to symmetry relations, i.e.:
\begin{equation}
\begin{split}
    \phi^{\nu_1}(x_1) \phi^{\nu_2}(x_2) & \neq (-1)^{\nu_1+\nu_2} \, \phi^{\nu_1}(x_1) \phi^{\nu_2}(x_2) \qquad \nu_1+ \nu_2 \,\,\, \text{odd},\\
    \qquad \phi^{\nu_1}(x_1) \phi^{\nu_2}(x_2) & \neq \phi^{\nu_1}(x_2) \phi^{\nu_2}(x_1);
\end{split}
\end{equation}
on the other hand, symmetries give non-trivial relations among integrals; they imply
\begin{equation}
\begin{split}
    I_{\nu_1 \nu_2}  = -I_{\nu_1 \nu_2} & \equiv 0 \qquad \nu_1+ \nu_2 \,\,\, \text{odd},\\
    I_{\nu_1 \nu_2} & \equiv  I_{\nu_2 \nu_1}.
\end{split}
\end{equation}

Employing systematically such kinds of symmetries~\cite{Gasparotto:2023roh} the number of independent integrals for $D$ dimensional lattices with $N=2^D$ points is summarized in the following table
\begin{equation}
    \begin{tabular}{ |c|c|c|c|c| } 
 \hline
 D & 1 & 2 & 3 & 4  \\ 
\text{$N$}& 2 & 4 & 8 & 16\\
\text{$\#$\,independent integrals}& 4 & 13 & 147 & 66\,524\\
 \hline
\end{tabular}
\label{eq:size_integrals}
\end{equation}
We appreciate the drastic reduction comparing~(\ref{eq.size_cohomology}) and~(\ref{eq:size_integrals}). 
\section{System of Differential Equations}
Having at our disposal a reduction algorithm and symmetry relations, we can embed the set of independent integrals into a vector
\begin{equation}
    \mathbf{I} = \left( I_1, \dots , I_{\text{$\#$\,indep.\,ints}} \right)^{\top},
\end{equation}
and, in principle, derive the corresponding system of first order DEQs with respect to each parameter that appears in the action
\begin{equation}
    \frac{d}{d  (\bullet)} \mathbf{I}(m^2, \lambda, t) =    A_{\bullet}(m^2, \lambda, t) \, \mathbf{I}(m^2, \lambda, t), \qquad \qquad \bullet=m^2, \lambda, t.
\label{eq:DEQ_system_All}
\end{equation}

While considering a system of differential equations such as the one in eq.~(\ref{eq:DEQ_system_All}), we face (at least) two problems: the (eventual) presence of singularities along the integration path and the determination of the boundary vector $\mathbf{I}_0$. It turns out that focusing on the system of DEQs with respect to the auxiliary parameter $t$ is particularly convenient. We will elucidate this claim in the next paragraph.\\

As we explained above, one of the features of our reduction algorithm is that it only produces polynomials in $t$. Therefore, the matrix controlling the system of DEQs with respect to the variable $t$ takes the following form
\begin{equation}
    A_t(m^2, \lambda, t) = \sum_{j=0}^{j_{\text{max}}} \mathcal{A}_j(m^2, \lambda) \, t^j,
\label{eq:At_DEQ}
\end{equation}
where $\mathcal{A}_j(m^2, \lambda)$ are sparse matrices with rational dependence on $(m^2, \lambda)$ and $j_{\text{max}}$ is a positive integer whose value is not important for the discussion\footnote{In the case of two lattice points in each direction$-$i.e. $L=2-$we have $j_{\text{max}}=2^D$, where $D$ is the number of dimensions.}. Eq.~(\ref{eq:At_DEQ}) has poles only at $t=\infty$, and it is therefore convenient to integrate the differential equations along the segment $t\in [0,1]$ where singularities are absent. The value $t=1$ is our target point (cf. eq.~(\ref{eq:Flow_action})), while at $t=0$ the lattice action drastically simplifies 
\begin{equation}
   S_{t=0,\bullet} =   0 \cdot S_{\bullet}^{\text{next neigh.}}+S_{\bullet}^{(2)}+\lambda S_{\bullet}^{(4)}, \qquad \bullet=E,M;
\end{equation}
there is no interaction among different lattice points, and $N$-dimensional integrals factorize into products of one-fold integrals (which can be expressed even in closed analytic form): the determination of the boundary vector $\mathbf{I}_0$ is therefore trivial.\\

Combining all the elements together, we can outline a computational strategy for the numerical evaluation of the integrals via differential equations$-$as the reader familiar with multi-loop calculus may recognize, our approach is inspired by the ``auxiliary mass flow" method~\cite{Liu:2017jxz} in the context of Feynman integrals$-$: 
\begin{itemize}
    \item Given input and fixed values $(\underline{m}^2, \underline{\lambda})$ compute the boundary vector $\mathbf{I}_0$;
    \item The matrix $A_{t}(\underline{m}^2, \underline{\lambda},t)$ is holomorphic in $\mathbb{C}$ (as a function of $t$), then also $\mathbf{I}(\underline{m}^2, \underline{\lambda},t)$ is holomorphic in $\mathbb{C}$ (as a function of $t$)~\cite{wasow1965asymptotic}; it takes the following form
    \begin{equation}
        \mathbf{I}(\underline{m}^2, \underline{\lambda},t) = \sum_{k=0}^{\infty} \mathbf{I}_{k}(\underline{m}^2, \underline{\lambda}) t^{k}
    \label{eq:ansatz_I}
    \end{equation}
    \item Plugging eq.~(\ref{eq:ansatz_I}) into the differential equation (cf. eq.~(\ref{eq:At_DEQ})), derive the recurrence relation
    \begin{equation}
        \mathbf{I}_k(\underline{m}^2, \underline{\lambda})= \frac{1}{k} \sum_{j=0}^{j_{\text{max}}} \mathcal{A}_j(\underline{m}^2, \underline{\lambda})\, \mathbf{I}_{k-j-1}(\underline{m}^2, \underline{\lambda});
    \label{eq:recursion}
    \end{equation}
    \item Truncate the recursion in eq.~(\ref{eq:recursion}) at the desired $k_{\text{max}}$; the corresponding expressions evaluated at $t=1$ yield the full set of integrals in the basis for the values $(\underline{m}^2, \underline{\lambda})$ given in input.
\end{itemize}
In summary, computing lattice correlation functions via a system of DEQs boils down to sparse matrix multiplication. For lattices of small size, such as the ones in~(\ref{eq.size_cohomology}) and~(\ref{eq:size_integrals}), the calculation can be performed with moderate computer resources: a desktop with $16$ GB RAM. 
\section{Results}
\begin{figure}
\begin{center}
\includegraphics[scale=0.4]{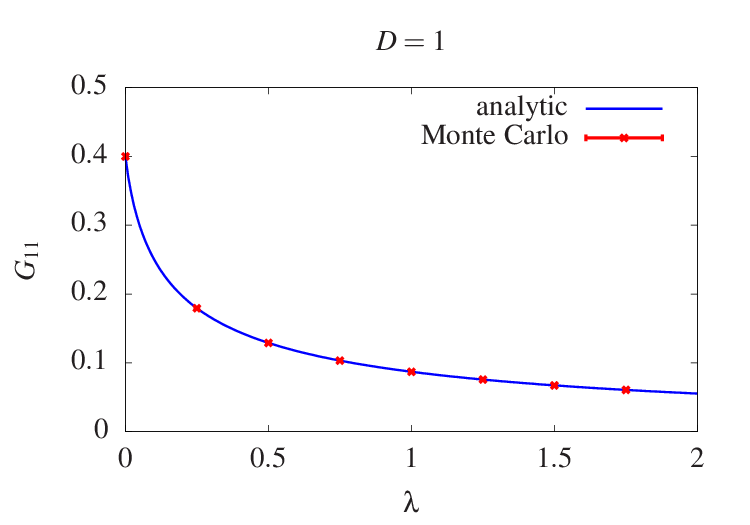}
\includegraphics[scale=0.4]{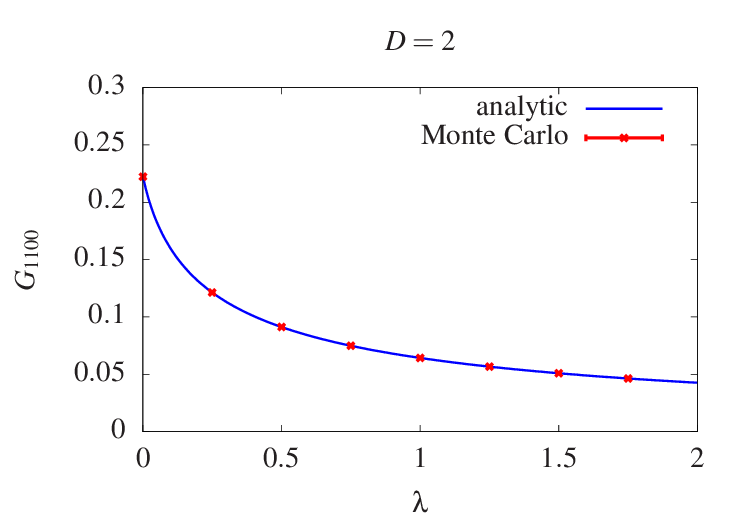}
\includegraphics[scale=0.4]{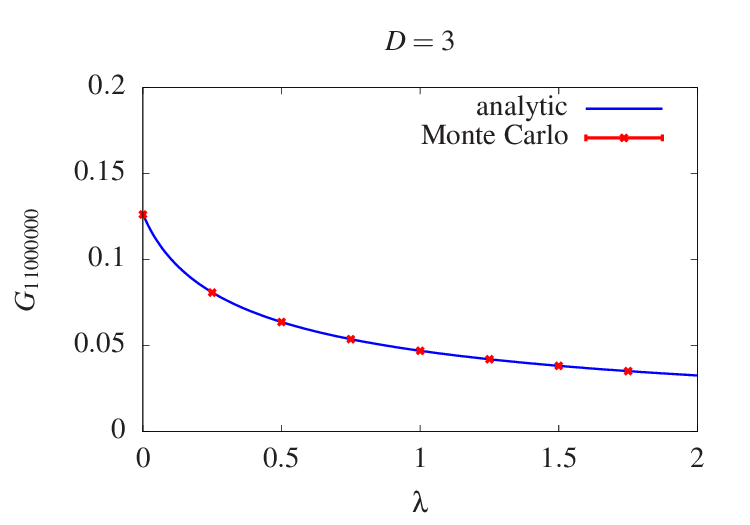}
\includegraphics[scale=0.4]{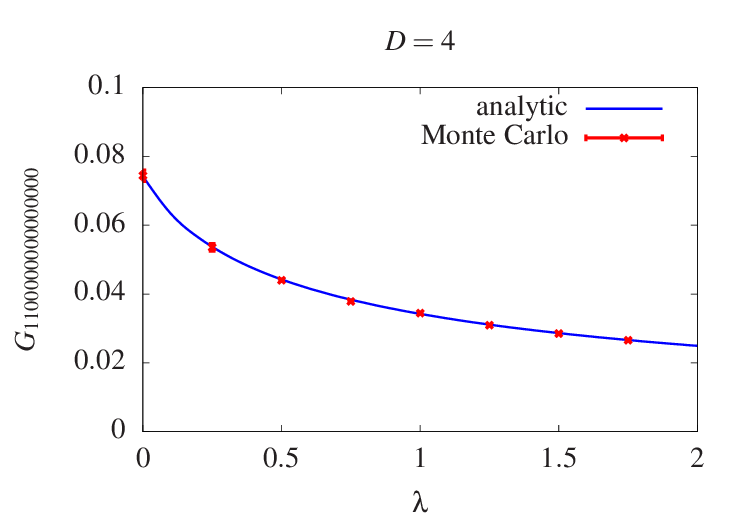}
\end{center}
\caption{
The Euclidean correlation function $G_{110\dots0}$ as a function of the coupling $\lambda$ (with $m^2\equiv\underline{m}^2=1$)
for $D=1,2,3,4$. The blue curve is obtained through the solution of the system of differential equations. Red dots are obtained via Monte-Carlo integration.
}
\label{fig_G11_euclidean}
\end{figure}

\begin{figure}
\begin{center}
\includegraphics[scale=0.4]{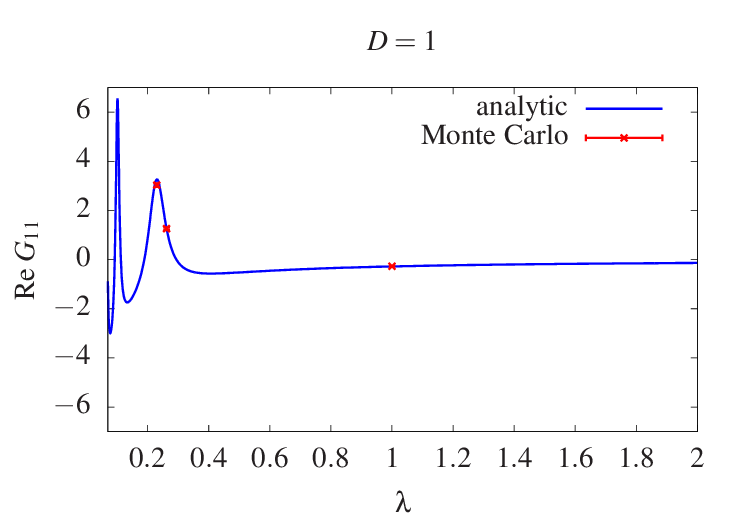}
\includegraphics[scale=0.4]{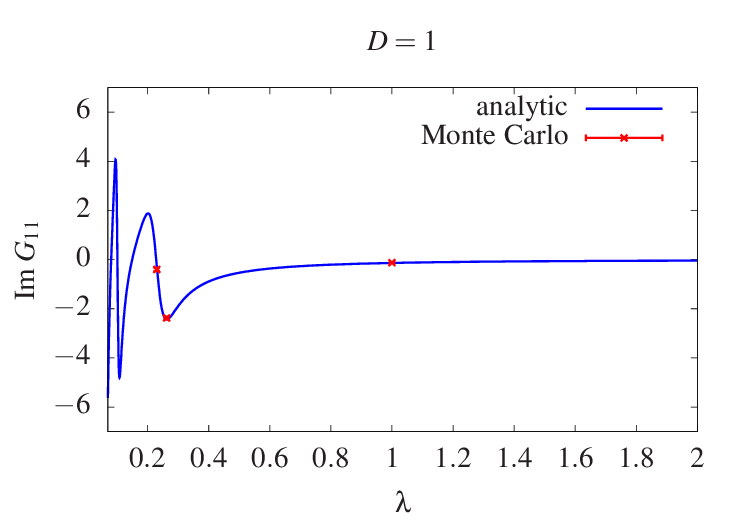}
\includegraphics[scale=0.4]{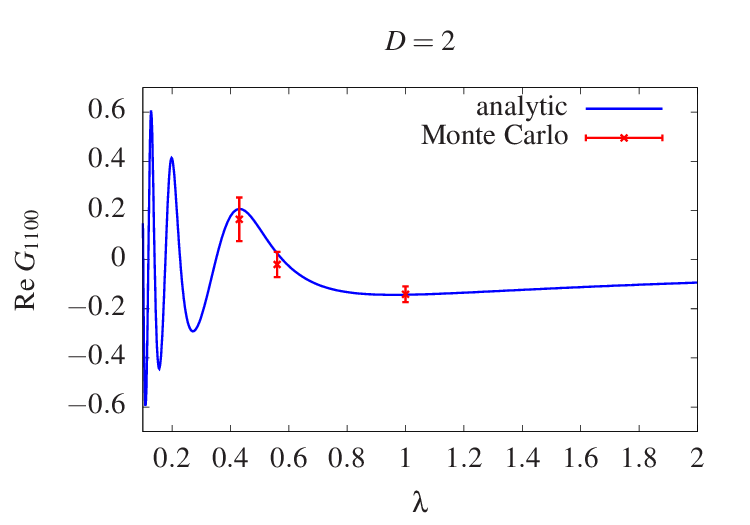}
\includegraphics[scale=0.4]{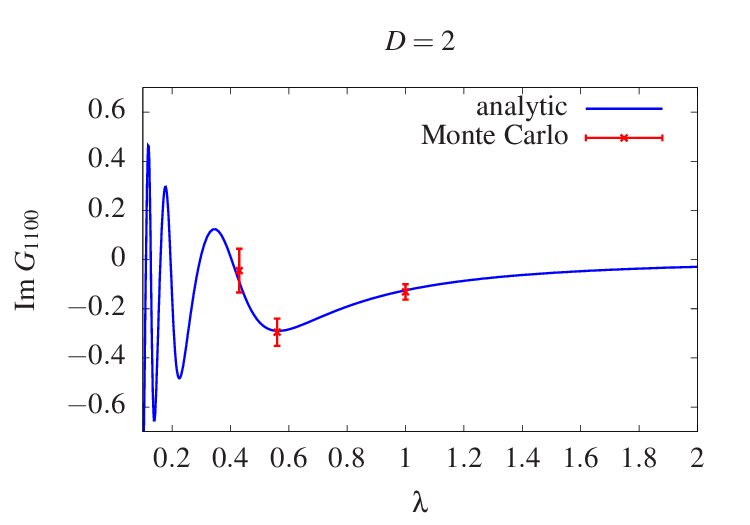}
\end{center}
\caption{
The Minkowskian correlation function $G_{110\dots0}$ in $D=1$ (upper panel) and $D=2$ (lower panel) as a function of the coupling $\lambda$ (with $m^2\equiv\underline{m}^2=1$).
The left plot shows the real part, the right plot shows the imaginary part. The values of $\lambda$ are in the range $\lambda \in [0.07,2]$ for $D=1$ and $\lambda \in [0.1,2]$ for $D=2$. The blue curve is obtained through the solution of the system of differential equations. Red dots are obtained via Monte-Carlo integration.
}
\label{fig_Minkowskian_D_eq_12}
\end{figure}

\begin{figure}
\begin{center}
\includegraphics[scale=0.4]{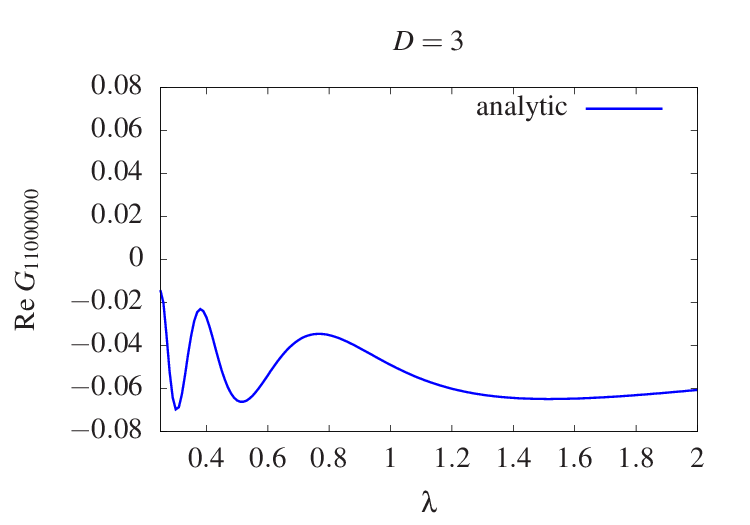}
\includegraphics[scale=0.4]{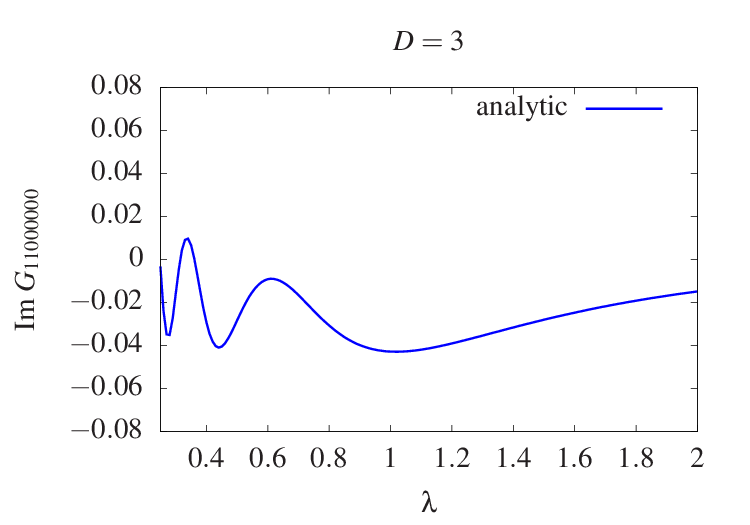}
\includegraphics[scale=0.4]{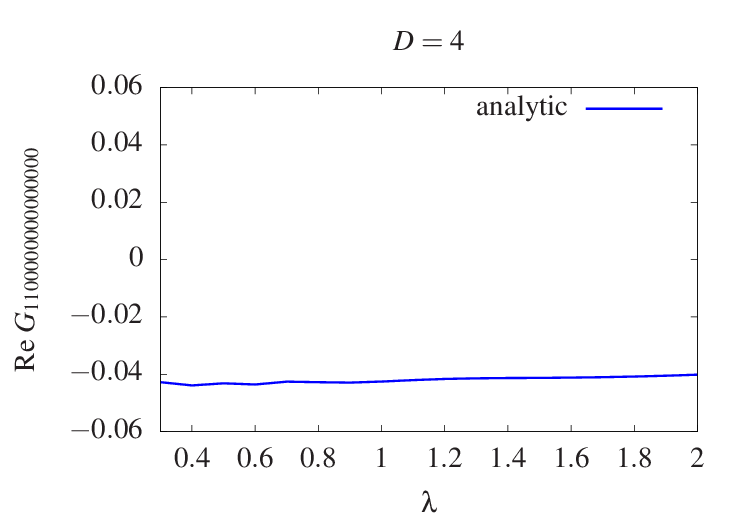}
\includegraphics[scale=0.4]{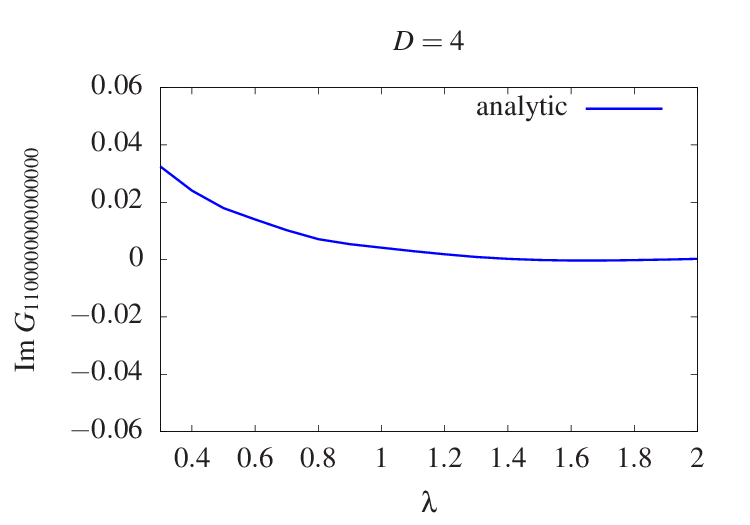}
\end{center}
\caption{
The Minkowskian correlation function $G_{110\dots0}$ in $D=3$ (upper panel) and $D=4$ (lower panel) as a function of the coupling $\lambda$ (with $m^2\equiv\underline{m}^2=1$).
The left plot shows the real part, the right plot shows the imaginary part. The values of $\lambda$ are in the range $\lambda \in [0.25,2]$ for $D=3$ and $\lambda \in [0.3,2]$ for $D=4$. The blue curve is obtained through the solution of the system of differential equations.
}
\label{fig_Minkowskian_D_eq_34}
\end{figure}

We consider lattices with $L=2$ points in each direction. We focus on the correlator
\begin{equation}
    G_{110 \dots,0} = \frac{I_{110 \dots 0}}{I_{0 0 0  \dots 0}},
    \label{eq:correlator_G11}
\end{equation}
meaning that we have one field at the origin and one field in the positive time-like direction. Correlation functions as a function of $\lambda$ for fixed $m^2\equiv\underline{m}^2=1$ in Euclidean signature for $D=1,2,3,4$ dimensions are presented in Figure~\ref{fig_G11_euclidean}. Eq.~(\ref{eq:correlator_G11}) as a function of $\lambda$ for fixed $m^2\equiv\underline{m}^2=1$ in Minkowskian signature for $D=1,2$ and $D=3,4$ dimensions are presented in Figure~\ref{fig_Minkowskian_D_eq_12} and in Figure~\ref{fig_Minkowskian_D_eq_34} respectively. Whenever possible, we compared our results against data obtained via Monte-Carlo integration, observing full agreement.
\section{Conclusions}
In this contribution, we discussed how methods developed in the context of perturbation theory can be applied to the study of lattice correlation functions at the non perturbative level. Thanks to IBPs identities$-$considered in the framework of twisted co-homology$-$and symmetry relations, it was possible to identify a minimal basis of integrals and, in principle, derive a system of DEQs for the above-mentioned basis with respect to each parameter that is present the action. We found it convenient to introduce, in a suitable way, an auxiliary quantity, dubbed auxiliary flow $t$ and consider the corresponding system of DEQs: this choice simplifies the solution and the determination of the boundary constants. Even if we considered lattices of small size with two points in each direction, our method is generally applicable to both Euclidean and Minkowskian signature.

Possible future directions may be oriented towards an optimization of our algorithm and setup (having in mind more robust computer resources) and applications beyond a scalar $\lambda \phi^4$ model: the addition of chemical potential or the study of Yang-Mills theory, to mention a few.
\section{Acknowledgements}
We would like to thank Michal Czakon, André Hoang, Arthur Lipstein, Pierpaolo Mastrolia, Paul McFadden, Harvey Meyer, Carlo Pagani, Georg von Hippel, Hartmut Wittig and Mao Zeng for interesting discussions and feedback. FG, SW and XX have been supported by the Cluster of Excellence Precision Physics, Fundamental Interactions, and Structure
of Matter (PRISMA EXC 2118/1) funded by the German
Research Foundation (DFG) within the German Excellence Strategy (Project ID 3908314).
\bibliographystyle{JHEP}
\bibliography{biblio}

\providecommand{\href}[2]{#2}\begingroup\raggedright\begin{thebibliography}{10}

\bibitem{Weinzierl:2020nhw}
S.~Weinzierl, {\it {Correlation functions on the lattice and twisted
  cocycles}},  {\em Phys. Lett. B} {\bf 805} (2020) 135449,
  [\href{http://arxiv.org/abs/2003.05839}{{\tt arXiv:2003.05839}}].

\bibitem{Gasparotto:2022mmp}
F.~Gasparotto, A.~Rapakoulias, and S.~Weinzierl, {\it {Nonperturbative
  computation of lattice correlation functions by differential equations}},
  {\em Phys. Rev. D} {\bf 107} (2023), no.~1 014502,
  [\href{http://arxiv.org/abs/2210.16052}{{\tt arXiv:2210.16052}}].

\bibitem{Gasparotto:2023roh}
F.~Gasparotto, S.~Weinzierl, and X.~Xu, {\it {Real time lattice correlation
  functions from differential equations}},  {\em JHEP} {\bf 06} (2023) 128,
  [\href{http://arxiv.org/abs/2305.05447}{{\tt arXiv:2305.05447}}].

\bibitem{Chetyrkin:1981qh}
K.~G. Chetyrkin and F.~V. Tkachov, {\it {Integration by Parts: The Algorithm to
  Calculate beta Functions in 4 Loops}},  {\em Nucl. Phys. B} {\bf 192} (1981)
  159--204.

\bibitem{Tkachov:1981wb}
F.~V. Tkachov, {\it {A Theorem on Analytical Calculability of Four Loop
  Renormalization Group Functions}},  {\em Phys. Lett. B} {\bf 100} (1981)
  65--68.

\bibitem{Laporta:2000dsw}
S.~Laporta, {\it {High precision calculation of multiloop Feynman integrals by
  difference equations}},  {\em Int. J. Mod. Phys. A} {\bf 15} (2000)
  5087--5159, [\href{http://arxiv.org/abs/hep-ph/0102033}{{\tt
  hep-ph/0102033}}].

\bibitem{Kotikov:1990kg}
A.~V. Kotikov, {\it {Differential equations method: New technique for massive
  Feynman diagrams calculation}},  {\em Phys. Lett. B} {\bf 254} (1991)
  158--164.

\bibitem{Remiddi:1997ny}
E.~Remiddi, {\it {Differential equations for Feynman graph amplitudes}},  {\em
  Nuovo Cim. A} {\bf 110} (1997) 1435--1452,
  [\href{http://arxiv.org/abs/hep-th/9711188}{{\tt hep-th/9711188}}].

\bibitem{Gehrmann:1999as}
T.~Gehrmann and E.~Remiddi, {\it {Differential equations for two loop four
  point functions}},  {\em Nucl. Phys. B} {\bf 580} (2000) 485--518,
  [\href{http://arxiv.org/abs/hep-ph/9912329}{{\tt hep-ph/9912329}}].

\bibitem{Mastrolia:2018uzb}
P.~Mastrolia and S.~Mizera, {\it {Feynman Integrals and Intersection Theory}},
  {\em JHEP} {\bf 02} (2019) 139, [\href{http://arxiv.org/abs/1810.03818}{{\tt
  arXiv:1810.03818}}].

\bibitem{Liu:2017jxz}
X.~Liu, Y.-Q. Ma, and C.-Y. Wang, {\it {A Systematic and Efficient Method to
  Compute Multi-loop Master Integrals}},  {\em Phys. Lett. B} {\bf 779} (2018)
  353--357, [\href{http://arxiv.org/abs/1711.09572}{{\tt arXiv:1711.09572}}].

\bibitem{wasow1965asymptotic}
W.~Wasow, {\em Asymptotic expansions for ordinary differential equations}.
\newblock Pure and Applied Mathematics, Vol. XIV. Interscience Publishers John
  Wiley \& Sons, Inc., New York-London-Sydney, 1965.

\end{thebibliography}\endgroup

\end{document}